\documentclass[
 ,secularistic%
,showpacs ,amssymb,nobibnotes, aps]{revtex4}
\usepackage{amsmath}
\usepackage{graphicx}
\input{epsf}

\usepackage{amsmath,amssymb}
\usepackage{bm}

\newcommand{\dalm}{\kern1pt\vbox{\hrule height 0.9pt\hbox{\vrule width 0.9pt\hskip 2.5pt\vbox{\vskip 5.5pt}\hskip 3pt\vrule width 0.3pt}\hrule height 0.3pt}\kern1pt}

\begin{document}



\title{Stellar Oscillations in Scalar-Tensor Theory of Gravity}
\author{Hajime Sotani}\email{sotani@gravity.phys.waseda.ac.jp}
\affiliation{Research Institute for Science and Engineering, Waseda University,\\
Okubo 3-4-1, Shinjuku, Tokyo 169-8555, Japan}

\author{Kostas D. Kokkotas}\email{kokkotas@auth.gr}
\affiliation{Department of Physics, Aristotle University of
Thessaloniki, Thessaloniki 54124, Greece.
}

\date{\today}

\begin{abstract}
We derive the perturbation equations for relativistic stars in
scalar-tensor theories of gravity and study the corresponding
oscillation spectrum. We show that the frequency of the emitted
gravitational waves is shifted proportionally to the scalar field
strength.  Scalar waves which might be produced from such
oscillations can be a unique probe for the theory, but their
detectability is questionable if the radiated energy is small.
However we show that there is no need for a direct observation of
scalar waves: the shift in the gravitational wave spectrum could
unambiguously signal the presence of a scalar field.
\end{abstract}

\maketitle
\section{Introduction}
\label{sec:Intro}

Scalar-tensor theories of gravity are an alternative or generalization
of Einstein's theory of gravity, where in addition to the tensor field
a scalar field is present. The theory has been proposed in its earlier
form about half century ago \cite{Fierz1956,Jordan1959,Brans1961}, and
it is a viable theory of gravity for a specific range of the coupling
strength of the scalar field to gravity
\cite{Damour1992,Will1993,Will2001}.  Actually, the existence of
scalar fields is crucial (e.g. in inflationary and quintessence
scenarios) to explain the accelerated expansion phases of the
universe. In addition, scalar-tensor theories of gravity can be
obtained from the low-energy limit of string theory and/or other gauge
theories. Experimentally, the existence of a scalar field has not yet
been probed, but a number of experiments in the weak field limit of
general relativity set severe limits on the existence and coupling
strengths of scalar fields \cite{Will2001,Esposito2004}.

A basic assumption is that the scalar and gravitational fields
$\varphi$ and $g_{\mu\nu}$ are coupled to matter via an ``effective
metric'' ${\tilde g}_{\mu\nu}= A^2(\varphi)g_{\mu\nu}$. The
Fierz-Jordan-Brans-Dicke \cite{Fierz1956,Jordan1959,Brans1961}
theory assumes that the ``coupling function'' has the form
$A(\varphi)=\alpha_0 \varphi$, i.e., it is characterized by a unique
free parameter $\alpha_0^2=(2\omega_{\rm BD}+3)^{-1}$, and all its
predictions differ from those of general relativity by quantities of
order $\alpha_0^2$ \cite{Damour1993}. Solar system experiments set
strict limits in the value of the Brans-Dicke parameter $\omega_{\rm
BD}$, i.e., $\omega_{\rm BD}\agt40000$, which suggests a very small
$\alpha_0^2<10^{-5}$ (see \cite{Bertotti2003,Esposito2004}).

In the early 1990s, based on a simplified version of scalar
tensor theory where $A(\varphi)=\alpha_0\varphi+\beta \varphi^2/2$,
Damour and Esposito-Farese \cite{Damour1993,Damour1996}
found that for certain values of the coupling parameter
$\beta$ the stellar models develop some strong field effects which
induce significant deviations from general relativity. This sudden
deviation from general relativity for specific values of the
coupling constants has been named ``spontaneous scalarization".
Harada \cite{Harada1998} studied in more detail models of
non-rotating neutron stars in the framework of the scalar-tensor
theory and he reported that ``spontaneous scalarization" is possible
for $\beta\alt-4.35$. DeDeo and Psaltis suggested
that the effects of scalar fields might be apparent in the
observed redshifted lines of X-rays and $\gamma$-rays observed
by Chandra and XMM-Newton \cite{DeDeo2003}
and in quasi-periodic oscillations (QPOs) \cite{DeDeo2004}.

Recently, we have examined the possibility to obtain the information
for the presence of the scalar field via gravitational wave
observations of oscillating neutron stars (\cite{SotaniKokkotas2004},
hereafter Paper I). This previous study has been done using the
so-called Cowling approximation. In this approximation one studies the
fluid oscillations freezing the perturbations of the spacetime and of
the scalar field. Even under these assumptions the effect of the
scalar field on the fluid perturbations can be significant. We showed
that for values of $\beta\alt-4.35$ the oscillation frequencies of the
fluid change drastically, and the observation of such oscillations can
verify or rule out the spontaneous scalarization phenomenon.

It has been suggested that stellar oscillations can provide a unique
tool for estimating the parameters of the star, i.e., mass, radius,
rotation rate, magnetic field and equation of state. These ideas
have been developed in the last decade in a series of papers
\cite{Andersson1996,Andersson1998,Benhar1999,Andersson2001a,Sotani2003,Sotani2004,Benhar2004},
where the properties of the various families of oscillation modes
have been used to probe the stellar parameters. The modes which are
mainly excited during the formation of a neutron star or during
starquakes and emit detectable gravitational waves are the fluid
$f$ and $p$ modes and the $w$ modes, which are associated to
oscillations of the spacetime \cite{Kokkotas2001}.

The effect of the scalar field on the $f$ and $p$ modes has been
examined in Paper I (in the Cowling approximation). In this article
we derive the full set of equations describing the oscillations of a
relativistic star, i.e., the perturbations of the fluid, the
spacetime, and the scalar field. Since the stellar models are
spherically symmetric, the oscillations can be classified as axial
or polar depending on their parity, and we can derive perturbation
equations for each class of perturbations. In the polar case we show
that the wave equations describing the perturbations of the fluid
and spacetime couple to the wave equation describing the
perturbations of the scalar field. In other words the polar
perturbations are affected not only by the presence of the scalar
field in the background but also by the coupling with the wave
equation describing the perturbations of the scalar field.  In the
axial case we find a single equation describing the perturbation of
the spacetime. This equation is not coupled to perturbations of the
fluid and of the scalar field: the scalar field only affects the
dynamics through its influence on the background.

The paper is organized as follows. In the next section we establish
our notation and briefly introduce the theoretical framework for the
scalar-tensor theories of gravity we consider. In Section
\ref{sec:III} we derive the perturbation equations which will be used
for the numerical estimation of the oscillation frequencies.  In
Section \ref{sec:IV} we describe the methods that have been used to
derive the axial $w$ modes in scalar-tensor theory of gravity and show
the results. In the final Section \ref{sec:V} we discuss the results
and their implications. We have included in 
three Appendices the
details of the various analytic and numerical calculations.
In Appendix \ref{sec:appendix_2} we describe the perturbations of the energy
momentum tensors for the fluid and the scalar field, while in the next
Appendix \ref{sec:appendix_3} we provide the analytic forms of the
perturbed Einstein equations. Finally, in the last Appendix
\ref{sec:appendix_4} we describe the numerical techniques that have
been used to calculated the quasinormal modes.  In this paper we adopt
the unit of $c=G=1$, where $c$ and $G$ denote the speed of light and
the gravitational constant, respectively, and the metric signature of
$(-,+,+,+)$.

\section{Stellar models in scalar-tensor theories of gravity}
\label{sec:II}

In this section we will study neutron star models in scalar-tensor
theory of gravity with one scalar field. This is a natural
extensions of Einstein's theory, in which gravity is mediated
not only by a second rank tensor (the metric tensor $g_{\mu\nu}$)
but also by a massless long-range scalar field $\varphi$. The
action is given by \cite{Damour1992}
\begin{equation}
 S = \frac{1}{16\pi G_*}\int\sqrt{-g_*}
   \left(R_*-2g_*^{\mu\nu}\varphi_{,\mu}\varphi_{,\nu}\right)d^4x
   + S_m\left[\Psi_m,A^2(\varphi)g_{*\mu\nu}\right],
\label{eq:action}
\end{equation}
where all quantities with asterisks are related to the ``Einstein
metric'' $g_{*\mu\nu}$, then $R_*$ is the curvature scalar for this
metric and $G_*$ is the bare gravitational coupling constant.
$\Psi_m$ represents collectively all matter fields, and $S_m$
denotes the  action of the matter represented by $\Psi_m$, which
is coupled to the ``Jordan-Fierz metric tensor''
$\tilde{g}_{\mu\nu}$. The field equations are usually written in terms of the
``Einstein metric'', but all non-gravitational experiments measure
the ``Jordan-Fierz'' or ``physical metric''. The ``Jordan-Fierz metric''
is related to the ``Einstein metric'' via the conformal
transformation,
\begin{equation}
 \tilde{g}_{\mu\nu} = A^2(\varphi)g_{*\mu\nu}.  \label{metric-relation}
\end{equation}
Hereafter, we denote by a tilde quantities in the ``physical frame''
and by an asterisk those in the  ``Einstein frame''. From the
variation of the action $S$ we get the field equations in the
Einstein frame
\begin{align}
 G_{*\mu\nu} &= 8\pi G_*T_{*\mu\nu}
              + T_{*\mu\nu}^{(\varphi)}\, , \label{basic1} \\
 \dalm_*\varphi &= -4\pi G_*\alpha(\varphi)T_*, \label{basic2}
\end{align}
where $T_{*\mu\nu}^{(\varphi)}$ is the energy-momentum of the massless scalar field, i.e.,
\begin{equation}
 T_{*\mu\nu}^{(\varphi)} \equiv 2 \left(\varphi_{,\mu} \varphi_{,\nu}
          - \frac{1}{2} g_{*\mu\nu} g_{*}^{\alpha \beta} \varphi_{,\alpha} \varphi_{,\beta}\right) \,
\end{equation}
and $T_*^{\mu\nu}$ is the energy-momentum tensor in the Einstein frame
which is related to the physical energy-momentum tensor $\tilde{T}_{\mu\nu}$ as follows,
\begin{equation}
 T_*^{\mu\nu} \equiv \frac{2}{\sqrt{-g_*}}\frac{\delta S_m}{\delta g_{*\mu\nu}}= A^6(\varphi)\tilde{T}^{\mu\nu} \, .
\end{equation}
The scalar quantities $T_{*}$ and $\alpha (\varphi)$ are defined as
$T_* \equiv T_{*\mu}^{\ \ \mu} = T_*^{\mu\nu}g_{*\mu\nu}$ and
$\alpha (\varphi) \equiv {d\ln A(\varphi)}/{d\varphi}$.
It is apparent that $\alpha(\varphi)$ is the only field-dependent
function which couples the scalar field with matter, for
$\alpha(\varphi)=0$ the theory reduces to general relativity.
Finally, the law of energy-momentum conservation
$\tilde{\nabla}_{\nu}\tilde{T}_{\mu}^{\ \nu}=0$ is transformed into
\begin{equation}
 \nabla_{*\nu}T_{*\mu}^{\ \ \nu} = \alpha (\varphi)T_*\nabla_{*\mu}\varphi, \label{basic3}
\end{equation}
and we set $\varphi_0$ as the cosmological value of the scalar field
at infinity. In this paper, we adopt the same form of conformal
factor $A(\varphi)$ as in Damour and Esposito-Farese
\cite{Damour1993}, which is
\begin{equation}
 A(\varphi) = e^{\frac{1}{2}\beta\varphi^2}, \label{A_varphi}
\end{equation}
i.e., $\alpha(\varphi)=\beta \varphi$ where $\beta$ is a real number.
In the case $\beta=0$ this scalar-tensor theory reduces to
general relativity, while ``spontaneous scalarization" occurs
for $\beta \alt -4.35$ \cite{Harada1998}.

We will model the neutron stars as self-gravitating perfect
fluid of cold degenerate matter in equilibrium. Then the metric describing an unperturbed,
non-rotating, spherically symmetric neutron star can be written as
\begin{equation}
 ds_*^2 = g_{*\mu\nu}dx^{\mu}dx^{\nu}
        = -e^{2\Phi}dt^2 + e^{2\Lambda}dr^2 + r^2(d\theta^2 + \sin^2\theta d\phi^2)
\end{equation}
where $e^{-2\Lambda} = 1-{2\mu(r)}/{r}$
while for the calculation of the mass function $\mu(r)$ and the
``potential'' function $\Phi(r)$ the reader should refer to Paper I.
Finally, the stellar matter is assumed to be a perfect fluid
\begin{equation}
 \tilde{T}_{\mu\nu} = \left(\tilde{\rho} +\tilde{P}\right)\tilde{U}_{\mu}\tilde{U}_{\nu}
                    + \tilde{P}\tilde{g}_{\mu\nu}.
\end{equation}
where $\tilde{U}_{\mu}$  is the four-velocity of the fluid,
$\tilde{\rho}$ is the total energy density, and
${\tilde P}$ is the pressure in the physical frame.



\section{Basic Perturbation Equations}
\label{sec:III}

In this section we present the equations describing perturbations of
the spacetime, scalar field, and fluid in a spherically symmetric
background. The equations we provide describe the non-radial
oscillations of spherically symmetric neutron stars in scalar-tensor
theories. We assume, in the physical frame, using the Regge-Wheeler
gauge \cite{RW1957}, the following form of the perturbed metric
tensor
\begin{equation}
\tilde{h}_{\mu\nu}=\tilde{h}_{\mu\nu}^{(-)}+\tilde{h}_{\mu\nu}^{(+)},
\end{equation}
where $\tilde{h}_{\mu\nu}^{(-)}$ denotes the {\em axial} (or {\em
odd parity}) part of metric perturbations
\begin{eqnarray}
\tilde{h}_{\mu\nu}^{(-)}&=&\sum_{l=2}^{\infty} \sum_{m=-l}^{l}\left(
 \begin{array}{cccc}
 0  &  0 & -h_{0,lm} {\sin^{-1}\theta} \partial_{\phi} & h_{0,lm} \sin\theta \, \partial_{\theta} \\
 0  &  0 & -h_{1,lm} {\sin^{-1}\theta} \partial_{\phi} & h_{1,lm} \sin\theta \, \partial_{\theta} \\
 \ast & \ast & 0 & 0 \\
 \ast & \ast & 0 & 0 \\
 \end{array}
 \right) Y_{lm},
\end{eqnarray}
and $\tilde{h}_{\mu\nu}^{(+)}$ denotes the {\em polar} (or {\em even
parity}) part of metric perturbations
\begin{eqnarray}
\tilde{h}_{\mu\nu}^{(+)}&=&\sum_{l=2}^{\infty} \sum_{m=-l}^{l}\left(
 \begin{array}{cccc}
 H_{0,lm} e^{2\Phi}  &  H_{1,lm}              & 0          & 0 \\
 *                   &  H_{2,lm} e^{2\Lambda} & 0          & 0 \\
 0                   &  0                     & r^2 K_{lm} & 0 \\
 0                   &  0                     & 0          & r^2 K_{lm} \sin^2 \theta \\
 \end{array}
 \right) Y_{lm} \, .
\end{eqnarray}
The functions $h_{0,lm}$, $h_{1,lm}$, $H_{0,lm}$, $H_{1,lm}$,
$H_{2,lm}$, and $K_{lm}$ describing the spacetime perturbations have
only radial and temporal dependence while
$Y_{lm}=Y_{lm}(\theta,\phi)$ is the spherical harmonic function.

Following the previous definitions the perturbed metric tensor
$h_{*\mu\nu}$ in the Einstein frame has the form:
\begin{align}
h_{*\mu\nu} =& \frac{1}{A^2} {\tilde h}_{\mu \nu} - \frac{2}{A}g_{*\mu\nu} \delta A \\
            =&\sum_{l=2}^{\infty} \sum_{m=-l}^{l} \frac{1}{A^2}\left(
 \begin{array}{cccc}
 \left(H_{0,lm} + 2A \delta A \right)e^{2\Phi} &  H_{1,lm}
 & -h_{0,lm} \sin^{-1} \theta \partial_{\phi}
 & h_{0,lm} \sin\theta \, \partial_{\theta} \\
 \ast & \left(H_{2,lm} - 2A \delta A \right) e^{2\Lambda}
 & -h_{1,lm} \sin^{-1} \theta \partial_{\phi}
 & h_{1,lm} \sin\theta \, \partial_{\theta} \\
 \ast & \ast & \left(K_{lm} - 2A \delta A\right) r^2 & 0 \\
 \ast & \ast & 0 & \left(K_{lm} - 2A \delta A\right) r^2 \sin^2\theta \\
 \end{array}
 \right) Y_{lm},
\label{Eq:h_Einstein}
\end{align}
where $\delta A \equiv A \beta \varphi \delta \varphi$ is the
perturbation of the conformal factor $A$ and $\delta \varphi$ is the
perturbation of the scalar field, where $\delta \varphi$ is a function
of $t$ and $r$ only.  The above definition of $h_{*\mu\nu}$ will be
used to derive the perturbation equations: in other words, we will
work out the perturbations in the Einstein frame and we will transform
back to the physical frame whenever we need it.

By defining the new set of perturbation functions $\hat{H}_0$,
$\hat{H}_1$, $\hat{H}_2$, $\hat{K}$, $\hat{h}_0$, and $\hat{h}_1$ as
follows
\begin{align}
 \hat{H}_{0,lm} &= \frac{1}{A^2}\left(H_{0,lm}+2A \delta A\right),
                & \hat{H}_{1,lm} &= \frac{1}{A^2}H_{1,lm}, \\
 \hat{H}_{2,lm} &= \frac{1}{A^2}\left(H_{2,lm}-2A \delta A\right),
                & \hat{K}_{lm} &= \frac{1}{A^2}\left(K_{lm}-2A \delta A\right), \\
 \hat{h}_{0,lm} &= \frac{1}{A^2}h_{0,lm}, & \hat{h}_{1,lm}
                &= \frac{1}{A^2}h_{1,lm}.
\end{align}
The perturbed metric $h_{*\mu\nu}$, in the Einstein frame, is
simplified considerably and reduced to the ``standard'' Regge-Wheeler
form of a perturbed spherical metric. We should notice that the scalar
perturbations $\delta A$ are linked only with the polar perturbation
functions $H_{0,lm}$, $H_{1,lm}$, $H_{2,lm}$, and $K_{lm}$. The axial
perturbation functions $h_{0,lm}$ and $h_{1,lm}$ are only affected by
the contribution of the scalar field to the background.

The perturbation equations will be derived by taking the variation
of Equations (\ref{basic1}) and (\ref{basic2})
\begin{align}
 \delta G_{*\mu\nu} &= 8\pi G_* \delta T_{*\mu\nu} + \delta T_{*\mu\nu}^{(\varphi)},
 \label{var_basic1} \\
 \dalm_* \delta \varphi &= -4\pi G_* \delta\left[\alpha(\varphi)T_*\right],
 \label{var_basic2}
\end{align}
The various components of $\delta T_{*\mu\nu}^{(\varphi)}$ are
expressed as linear combinations of $\delta \varphi$ and ${\tilde
h}_{\mu\nu}$. In the Einstein frame, the perturbed energy-momentum
tensor describing the matter fields $\delta T_{*\mu\nu}$, is some
linear combination of the velocity variation of the fluid $\delta
{\tilde U}^i\sim\left(W Y_{lm},V\partial_\theta
Y_{lm}-u\sin^{-1}\theta \partial_\phi Y_{lm},V\partial_\phi
Y_{lm}+u\sin \theta \partial_\theta Y_{lm} \right)$, the density and
pressure variations ($\delta \tilde{\rho}$ and $\delta \tilde{P}$)
together with the variation of the scalar field ($\delta \varphi$ or
$\delta A$) and the metric perturbation $h_{*\mu\nu}$. The explicit
form of the energy-momentum tensor is given in Appendix
\ref{sec:appendix_2}.

The linearized Einstein equations (\ref{var_basic1}) will be
written as follows. From the $tt$, $tr$, $rr$ components and the sum
of the components $\theta\theta$ and $\phi\phi$, we get
\begin{equation}
 \sum_{l,m} A_{lm}^{(I)} Y_{lm} = 0\ \ (I=0\ \mbox{to}\ 3), \label{perturbed-einstein1}
\end{equation}
where the four expressions $A_{lm}^{(I)}=0$ are given in Appendix
\ref{sec:appendix_3}. They contain combinations of $\hat{H}_0$,
$\hat{H}_1$, $\hat{H}_2$, $\hat{K}$, $W$, $\delta \tilde{P}$,
$\delta \tilde{\rho}$, $\delta \varphi$ and their temporal and
spatial derivatives. It is worth noticing that all four equations
above are descibing only polar perturbations. In a similar way, from
the $t\theta$, $t\phi$, $r\theta$, and $r\phi$ components, we get
four more equations
\begin{gather}
  \sum_{l,m} \left\{\alpha_{lm}^{(J)} \partial_{\theta} Y_{lm}
                + \beta_{lm}^{(J)}\frac{1}{\sin\theta} \partial_{\phi} Y_{lm} \right\} = 0\ \ (J=0,1),
                  \label{perturbed-einstein2} \\
  \sum_{l,m} \left\{\beta_{lm}^{(J)} \partial_{\theta} Y_{lm}
                - \alpha_{lm}^{(J)}\frac{1}{\sin\theta} \partial_{\phi} Y_{lm} \right\} = 0\ \ (J=0,1)
                  \label{perturbed-einstein3}.
\end{gather}
Here the expressions $\alpha_{lm}^{(J)}$ are some linear
combinations of polar perturbation functions while on the other
hand, the expression $\beta_{lm}^{(J)}$ is a combination of only
axial perturbation functions (see Appendix \ref{sec:appendix_3}).

Furthermore, from the $\theta\phi$ component and the subtraction of
$\theta\theta$ and $\phi\phi$ components, we get two more equations
\begin{gather}
 \sum_{l,m} \left\{ s_{lm} X_{lm} - t_{lm} \sin \theta W_{lm}\right\} =0, \label{perturbed-einstein4} \\
 \sum_{l,m} \left\{ t_{lm} X_{lm} + s_{lm} \sin \theta W_{lm}\right\} =0, \label{perturbed-einstein5}
\end{gather}
where $s_{lm}$ and $t_{lm}$ describe polar and axial type
perturbations respectively while $X_{lm}$ and $W_{lm}$ are defined
as
\begin{equation}
 X_{lm} = 2 \partial_{\phi} \left(\partial_{\theta} - \frac{\cos \theta}{\sin \theta}\right)
 Y_{lm} \quad \mbox{and} \quad
 W_{lm} = \left(\partial_{\theta}^2 - \frac{\cos \theta}{\sin \theta} \partial_{\theta}
         - \frac{1}{\sin^2 \theta} \partial_{\phi}^2 \right) Y_{lm}.
\end{equation}

Taking the product of Equations (\ref{perturbed-einstein1}) --
(\ref{perturbed-einstein5}) with $\bar{Y}_{lm}$, integrating over the
solid angle and looking at components with fixed values of $l$ and
$m$, we get ten partial differential equations in the variables $t$
and $r$:
\begin{gather}
 A_{lm}^{(I)}=0 \, , \qquad  \alpha_{lm}^{(J)} =0 \, , \qquad s_{lm} = 0, \quad
 (I=0 \, \, - \, \, 3 \, \mbox{and} \, J=0,1) \label{polar-perturbed-einstein} \\
 \beta_{lm}^{(J)} =0 \, , \qquad t_{lm} = 0 \quad (J=0,1). \label{axial-perturbed-einstein}
\end{gather}
Equations (\ref{polar-perturbed-einstein}) describe the polar
perturbations, and Equations (\ref{axial-perturbed-einstein}) describe
the axial perturbations. The analytic expressions for Equations
(\ref{axial-perturbed-einstein}), i.e., Eqs. (\ref{beta0_lm}),
(\ref{beta1_lm}) and (\ref{t_lm}), do not couple to the perturbations
of the scalar field $\delta A$ or $\delta \varphi$.  Therefore, the
perturbed scalar field is coupled only to the polar perturbations.

\subsection{Axial perturbations}
It is quite easy to derive a wave equations for the axial perturbations by combining equations (\ref{beta1_lm}) and (\ref{t_lm})
\begin{align}
 \ddot{X} &- e^{\Phi - \Lambda} \left( e^{\Phi - \Lambda} X'\right)'
      +  e^{2\Phi} \left(\frac{l(l+1)}{r^2} - \frac{6\mu}{r^3}
      + 4\pi G_* \left(\tilde{\rho}-\tilde{P}\right)A^4\right)X =0, \label{axial_wave}
\end{align}
where we introduce the new function $X(t,r)$ defined as $\hat{h}_1 =
e^{\Lambda-\Phi}X r$.
%
This equation does not couple to the scalar field perturbations, as
we have mentioned earlier, and the effects of the scaler field will
enter only via the background terms. Thus for $\beta=0$, i.e., $A=1$,
it reduces to the standard wave equation describing axial
perturbations \cite{CF91}. Finally, from equation (\ref{beta1_lm})
we get the following relation,
\begin{equation}
 \dot{u} = -e^{-2\Phi} \dot{\hat{h}}_0, \label{dot-u}
\end{equation}
which suggests that there are no axial oscillatory fluid motions,
i.e., they have zero frequency and represent stationary currents.
Thus axial perturbations are described only by a single wave
equation (\ref{axial_wave}) which does not couple  neither to polar
fluid or spacetime perturbations nor to the perturbed scalar field
and it can be studied independently.

\subsection{Polar perturbations}
The equations describing polar perturbations can be simplified
introducing a new set of perturbation functions.  Introducing these
functions we can reformulate the seven equations describing polar
perturbations as a pair of wave equations and a constraint equation,
using a procedure similar to Ref.~\cite{Allen98}. The new metric
perturbation functions will be
\begin{equation}
 F(t,r) = r \hat{K}, \quad \mbox{and} \quad
 S(t,r) = \frac{e^{2\Phi}}{r} \left(\hat{H}_0 - \hat{K}\right),
\end{equation}
while the fluid perturbations can be described by variation of the enthalpy function, i.e.,
\begin{align}
 H(t,r) &= \frac{\delta \tilde{P}}{\tilde{\rho} + \tilde{P}}.
\end{align}
The system of equations describing the polar perturbations can be
reduced to the following pair of wave equations;
\begin{align}
 \ddot{F} &- e^{2(\Phi - \Lambda)} F'' - e^{2\Phi} \left[4\pi G_* r \left(\tilde{P} - \tilde{\rho}\right)A^4
      + \frac{2\mu}{r^2}\right] F'
      + 2r \left(\tilde{\rho} + \tilde{P}\right) \left(1-\frac{1}{C_s^2}\right) A^4 e^{2\Phi} H \nonumber \\
      &+ e^{2\Phi} \left[\frac{l(l+1)}{r^2} - \frac{2\mu}{r^3}
      - 4\pi G_* \left(3\tilde{\rho} + \tilde{P}\right) A^4 - 2 e^{-2\Lambda} \Psi^2\right] F \nonumber \\
      &+ 2 e^{-2\Lambda} \left[1-r^2 \Psi^2
      - 4\pi G_* r^2 e^{-2\Lambda} \left(\tilde{\rho} + \tilde{P}\right) A^4\right] S
      + 8 e^{2\Phi} \left[\Psi e^{-2\Lambda}
      + 4\pi G_* \alpha r \left(\tilde{P} - \tilde{\rho}\right) A^4\right] \delta \varphi = 0, \label{ddot_F}
\end{align}
and
\begin{align}
 \ddot{S} &- e^{2(\Phi - \Lambda)} S'' - e^{2\Phi} \left[4\pi G_* r \left(\tilde{P} - \tilde{\rho}\right)A^4
      + \frac{2\mu}{r^2}\right] S'
      + e^{2\Phi} \left[\frac{l(l+1)}{r^2} - \frac{2\mu}{r^3}
      - 4\pi G_* \left(3\tilde{P} + \tilde{\rho}\right) A^4 + 2e^{-2\Lambda} \Psi^2\right] S \nonumber \\
      &+ \frac{4e^{4\Phi}}{r^5} \left\{{\Phi'}^2 r^3 e^{-2\Lambda} + 4\pi G_* \tilde{\rho} A r^3
      - 3\mu + \frac{4\pi G_*}{r} \left(\tilde{\rho} - 3\tilde{P}\right) \alpha A^4 \Psi
      + \left[ 4\pi G_* \left(\tilde{\rho} - \tilde{P}\right) r^5 A^4 - 2r^3 \right] \Psi^2\right\} F \nonumber \\
      &+ 4 e^{4\Phi} \left\{ \Psi^3 e^{-2\Lambda}
      + \left[ 8\pi G_* \left(2\tilde{P} - \tilde{\rho}\right)A^4
      + \frac{10\mu}{r^3} - \frac{2}{r^2}\right] \Psi\right\} \delta \varphi = 0 \, . \label{ddot_S}
\end{align}
From the $tt$ component of the perturbed Einstein equations we get the
Hamiltonian constraint:
\begin{align}
 F'' &+ \left[\frac{e^{2\Lambda}}{r^2} \left(\mu - 4\pi G_* r^3 \tilde{\rho} A^4 \right)
      - \frac{1}{2} r \Psi^2 \right] F'
      + \frac{e^{2\Lambda}}{r^3} \left[ 12\pi G_* \tilde{\rho} r^3 A^4 - \mu - rl(l+1)
      + \frac{1}{2} r^3 \Psi^2 e^{-2\Lambda} \right] F \nonumber \\
      &- r e^{-2\Phi} S'
      + e^{-2\Phi + 2\Lambda} \left[ \Psi^2 r^2 e^{-2\Lambda}
      + 8 \pi G_* r^2 \left(\tilde{\rho} + \tilde{P}\right) A^4 - 2 e^{-2\Lambda} - \frac{l(l+1)}{2}\right]S
        \nonumber \\
      &+ \frac{8\pi G_* r}{C_s^2} \left( \tilde{\rho} + \tilde{P} \right) e^{2\Lambda} A^4 H
      + 2r \Psi \delta \varphi' + 32\pi G_* r A^4 e^{2\Lambda} \tilde{\rho} \alpha \delta \varphi = 0. \label{constraint}
\end{align}

Furthermore, for the special form of the conformal factor,
i.e., $A=\exp(\beta\varphi^2/2)$ or $\alpha=\beta \varphi$,
from Equation (\ref{var_basic2}) we obtain a wave
equation for the perturbed scalar field $\delta \varphi$:
\begin{align}
  \delta \ddot{\varphi} -& e^{\Phi-\Lambda} \left(e^{\Phi - \Lambda} \delta \varphi'\right)'
      - \frac{2}{r} e^{2\Phi - 2\Lambda} \delta \varphi' + e^{2\Phi} \left[\frac{l(l+1)}{r^2}
      + 4 e^{-2\Lambda} \Psi^2
      - 4 \pi G_* A^4 \left(\tilde{\rho} - 3\tilde{P}\right)
        \left(4 \alpha^2 + \beta \right)\right] \delta \varphi
        \nonumber \\
      =& \left[r^2 e^{-2\Lambda} \Psi^3 + \frac{2 \mu \Psi}{r} + 4 \pi G_* r A^4
        \left\{2 r \Psi \tilde{P} - \alpha \left(\tilde{\rho} - 3 \tilde{P}\right)\right\}\right] S
        \nonumber \\
      &+ e^{2 \Phi} \left[e^{-2 \Lambda} \Psi^3 + \frac{2 \mu \Psi}{r^3} + 4 \pi G_* A^4
        \left\{2 \Psi \tilde{P} - \frac{\alpha}{r}\left(\tilde{\rho} - 3 \tilde{P}\right)\right\}\right]F
       - 4 \pi G_* A^4 \alpha e^{2 \Phi} \left(\tilde{\rho} + \tilde{P}\right)\left(\frac{1}{C_s^2} - 3\right)H,
      \label{perturbation_scalarphi}
\end{align}
where $C_s^2 = d\tilde{P}/d\tilde{\rho}$.

Finally, by linearizing equation (\ref{basic3}), i.e., the
energy-momentum conservation equations, we get a system of evolution
equations for the components of the perturbed velocity and the
density perturbation

\begin{align}
 &W' + \frac{1}{C_s^2} e^{2\Lambda-2\Phi}  \dot{H} + e^{2\Lambda - 2\Phi}
 \left( 3 \alpha \delta \dot{\varphi}
     + \frac{r}{2} e^{- 2\Phi} \dot{S} + \frac{3}{2r}  \dot{F}
     \right)
     \nonumber \\
     &\hspace{2cm}- \frac{l(l+1)}{r^2} e^{2\Lambda} V + \left[2\Phi' - \Lambda' + \frac{2}{r} + 3 \alpha \Psi
     - \frac{1}{C_s^2} \left(\Phi' + \alpha \Psi \right)\right] W = 0,  \label{perturbation_energymomuntum1}\\
 &\dot{W} + H' + \alpha \delta \varphi' + \frac{r}{2} e^{-2\Phi} S'
 - \frac{1}{2r} F'
     + e^{-2\Phi} \left(\frac{1}{2} - r \Phi' + r^2 \Psi^2 + \frac{2 \mu}{r} e^{2\Lambda}
     + 8 \pi G_* r^2 A^4 e^{2\Lambda} \tilde{P} \right) S \nonumber \\
     &\hspace{2cm}+ \left(\Psi^2 + \frac{2 \mu}{r^3} e^{2\Lambda} + \frac{1}{2r^2}
     + 8 \pi G_* A^4 e^{2\Lambda} \tilde{P}\right) F
     + (\beta - 4) \Psi \delta \varphi = 0, \label{perturbation_energymomuntum2}\\
 &\dot{V} - \frac{r}{2} e^{-2\Phi} S - \frac{1}{2r} F + H + \alpha \delta \varphi = 0.
\label{perturbation_energymomuntum3}
\end{align}

Concluding, the dynamics of the polar perturbation is described by
the system of evolution equations (\ref{ddot_F}), (\ref{ddot_S}) and
(\ref{perturbation_scalarphi}) together with the constraint equation
(\ref{constraint}). The rest of the functions describing the
spacetime and fluid perturbations will be computed by taking proper
combinations of $F$, $S$, $H$ and $\delta \varphi$.

\section{Spacetime Perturbations in Scalar-Tensor Gravity}
\label{sec:IV}

In Paper I we have studied stellar perturbations in scalar-tensor
theories of gravity freezing the spacetime and scalar field
perturbations. This is the so-called Cowling approximation, in which
we only consider perturbations of the fluid. In practice we worked
with a system of equations similar to
(\ref{perturbation_energymomuntum1}) -
(\ref{perturbation_energymomuntum3}), but setting
$H_1=H_0=H_2=K=\delta \varphi =0$. Even under this approximation,
spontaneous scalarization has a remarkable effect on the oscillation
spectra of the $f$ and $p$ modes. Based on this observation we
proposed in Paper I that a successful detection of gravitational
waves from oscillating stars will provide us with a tool to
constrain the phenomenon of 'spontaneous scalarization'.

The quasinormal modes of the fluid perturbations described in Paper
I will be affected by the coupling to the spacetime and scalar field
perturbations. This is an interesting problem on its own. However,
since we have already shown that the effect of spontaneous
scalarization is quite strong when we limit consideration to
perturbations of the fluid, in this paper we will examine the effect
of the scalar field on the quasinormal modes describing the pure
spacetime oscillations, i.e., the so-called $w$ modes
\cite{Kokkotas1992}. The $w$ modes are similar to quasinormal modes
of black holes. They have higher frequencies and shorter damping
times that the $f$ modes, typical frequencies being of order $7-12$
kHz and damping times of order of 0.1~ms. These properties of the
$w$ modes are common to axial and polar perturbations
\cite{Benhar1999}. In the case of polar perturbations the $w$ modes
are associated to small fluid motions, while in the axial case there
is no coupling with the fluid at all. This is the reason why here we
choose to study the effect of the scalar field considering only the
axial perturbations, described by the single wave equation
(\ref{axial_wave}). We expect the effect of the scalar field on the
axial and polar $w$ modes to be of the same order of magnitude. It
should be pointed out here that, according to recent collapse
calculations \cite{Baiotti05}, the $w$ modes are significantly
excited. This adds further motivation to our study of the effects of
scalar fields on $w$ mode oscillations.

The equations needed to construct the equilibrium stellar
configurations as well as the equations of state (EOS) used are
described in Paper I.  In Paper I we also considered cases where the
asymptotic value of the scalar field $\varphi_0 \ne 0$, here, for
simplicity, we only deal with scalar fields with $\varphi_0=0$.

To compute the quasinormal frequencies of the axial $w$ modes we will
use two different techniques. In the first approach we carry out time
evolutions of Equation (\ref{axial_wave}) and Fourier transform the
signal at infinity; in the second approach we assume a harmonic time
dependence of the perturbations, and the corresponding boundary value
problem.

\subsection{Evolving the axial perturbation equation}

The time evolution of the 1+1 equation (\ref{axial_wave}) is rather
simple to obtain. We set some arbitrary initial data (for example a
Gaussian pulse) in equation (\ref{axial_wave}) and evolve these data
for some time. Then we compute the oscillation frequencies by taking
the Fourier transform of the signal emitted at infinity.

\begin{center}
\begin{figure}[htbp]
\includegraphics[height=6cm]{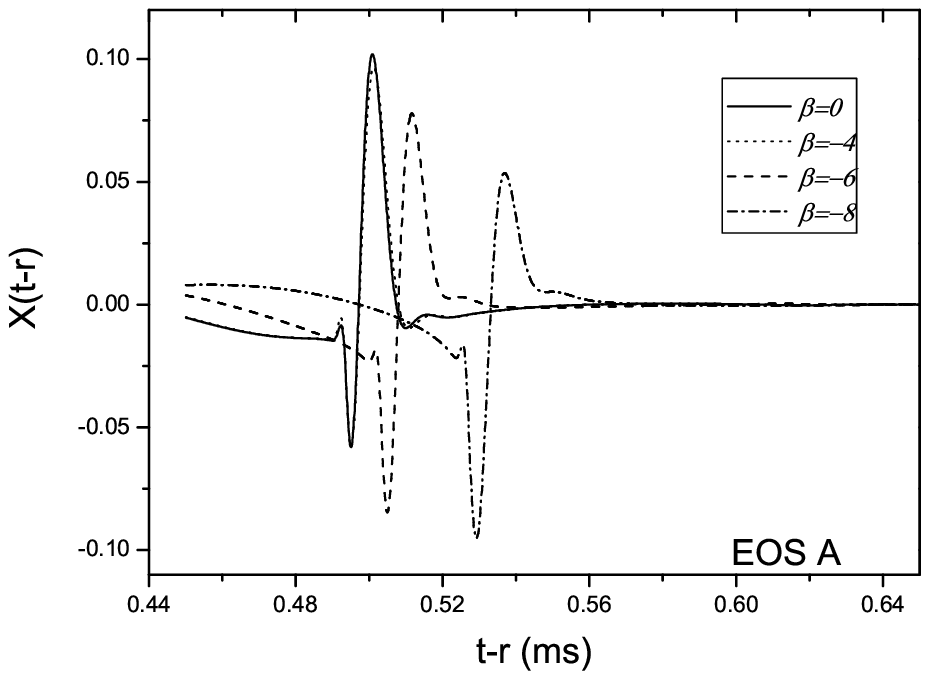}
\includegraphics[height=6cm]{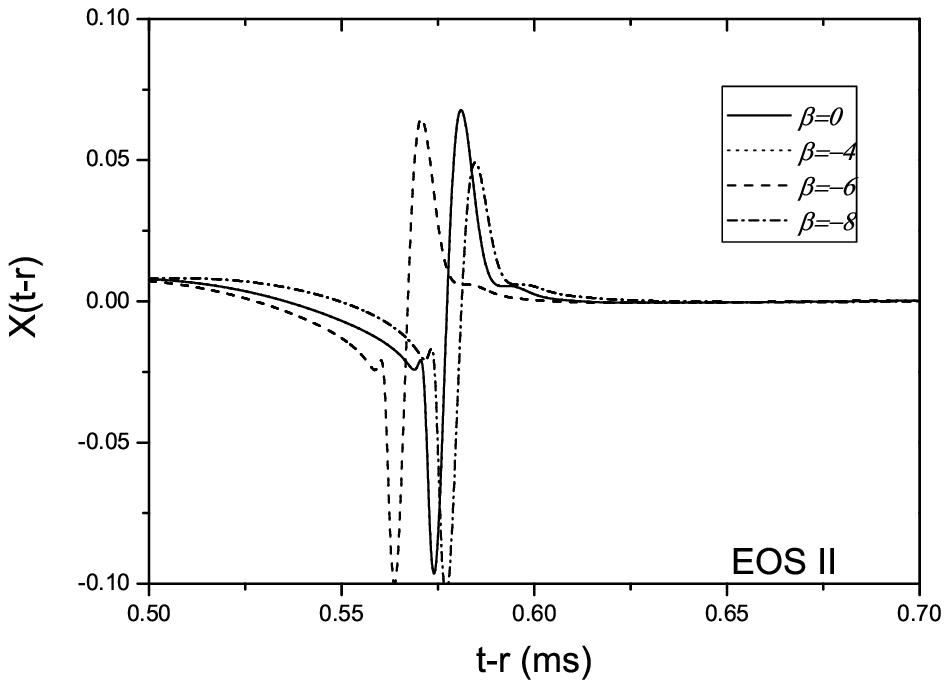} \caption{
Waveforms describing the evolution of $X$ for EOS A (left panel) and
EOS II (right panel) generated for stellar models with
$M_{ADM}=1.4M_{\odot}$ for $\beta=0$, $-4$, $-6$ and $-8$. The cases
$\beta=0$ and $\beta=-4$ are indistinguishable.}  \label{fig_wave}
\end{figure}
\end{center}

In Figure \ref{fig_wave} we show the waveforms observed at a distance
of about 300km from a neutron star with Arnowit-Deser-Misner (ADM) mass $1.4M_{\odot}$.
It is noticable that the arrival time of wave for different values
of $\beta$ is not the same, because the effective potential due to
central neutron star changes as a function of $\beta$.  In this
figure, we can see that the waveforms for $\beta=0$ and for
$\beta=-4$ are identical. This result can be understood as follows.
For the axial perturbation, the gravitational wave is not coupled
with the perturbation of the matter and the scalar field. So the
presence of the scalar field is realized only due to the modified
background. On the other hand, with $\varphi_0=0$, the central value
of scalar field $\varphi_c$ is zero for any $\beta>-4.35$ (see
\cite{SotaniKokkotas2004}). Thus for values of $\beta>-4.35$ the
effect of the scalar field is insignificant and it will affect the
results only for $\beta<-4.35$.
%
%

\subsection{Boundary value method}

Our second approach to calculate the quasinormal frequencies of the
axial $w$ modes is more involved than simple time evolutions.
However, using time evolutions we can only identify those modes that
are significantly excited by a certain set of initial data. For
example, using time evolutions it is not easy to identify
quasinormal modes that damp out very fast.  The approach we present
here allows us to calculate both slowly and highly damped
quasinormal modes. We Fourier-expand the wave equation
(\ref{axial_wave}) as $X(t,r)=X(r)\exp (i\omega t)$ and get
\begin{align}
 X'' + \frac{2\mu}{r(r-2\mu)} X'
     + \left(1-\frac{2\mu}{r}\right)^{-1} \left[\omega^2 e^{-2\Phi}
     - \frac{l(l+1)}{r^2} + \frac{6\mu}{r^3}
     - 4\pi G_* \left(\tilde{\rho} - \tilde{P} \right) A^4 \right] X = 0.
     \label{wave-equation}
\end{align}
In this way we obtain an eigenvalue problem: the complex quasinormal
modes $\omega$ can be obtained imposing appropriate boundary
conditions.  In our case, the boundary conditions are that $X(r)$
should be regular at the stellar center and that there are no incoming
waves at infinity. Inside the star we can just integrate the above
differential equation; outside the star we use appropriate asymptotic
expansions to ensure that there is no incoming radiation. Here we
adopt a variant of Leaver's continued fraction method
\cite{Leaver1985}, that has been originally used for the calculation
of quasinormal modes of black holes.  The procedure is described in
detail in Appendix \ref{sec:appendix_4}.
\begin{center}
\begin{figure}[htbp]
\includegraphics[height=6cm]{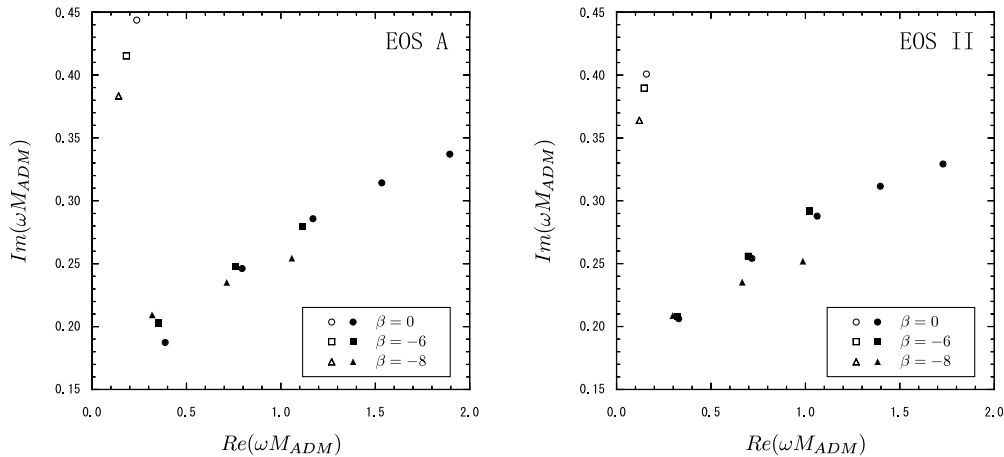}
\caption{The frequencies of the axial $w$ modes for $l=2$, here we
have considered stellar models with $M_{ADM}=1.4M_{\odot}$. The left
panel corresponds to EOS A and the right panel to EOS II. We show
both the ``ordinary'' $w$ modes ($w_1$, $w_2$, $w_3$, ...) and the
$w_{\rm II}$ modes (upper left corner in the plot). Open diamonds,
squares and triangles correspond to the $w_{\rm II}$ modes with
$\beta=0$, $\beta=-6$ and $\beta=-8$, respectively. Similarly,
filled symbols refer to the fundamental $w$ mode and its
overtones.}
\label{fig_m140}
\end{figure}
\end{center}
%
%

\subsection{Results}

\begin{center}
\begin{figure}[htbp]
\includegraphics[height=6cm]{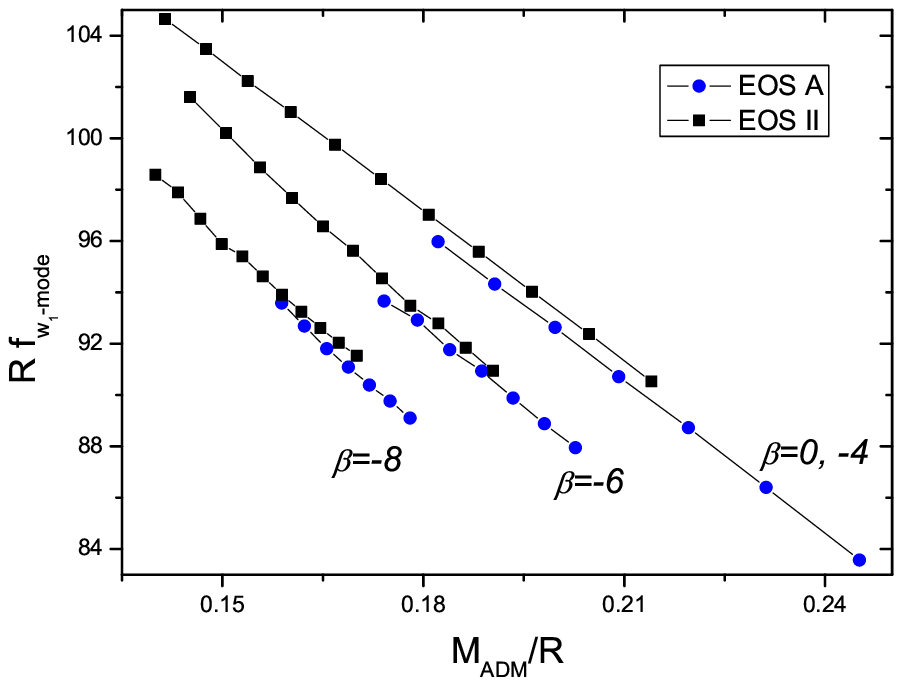}
\includegraphics[height=6cm]{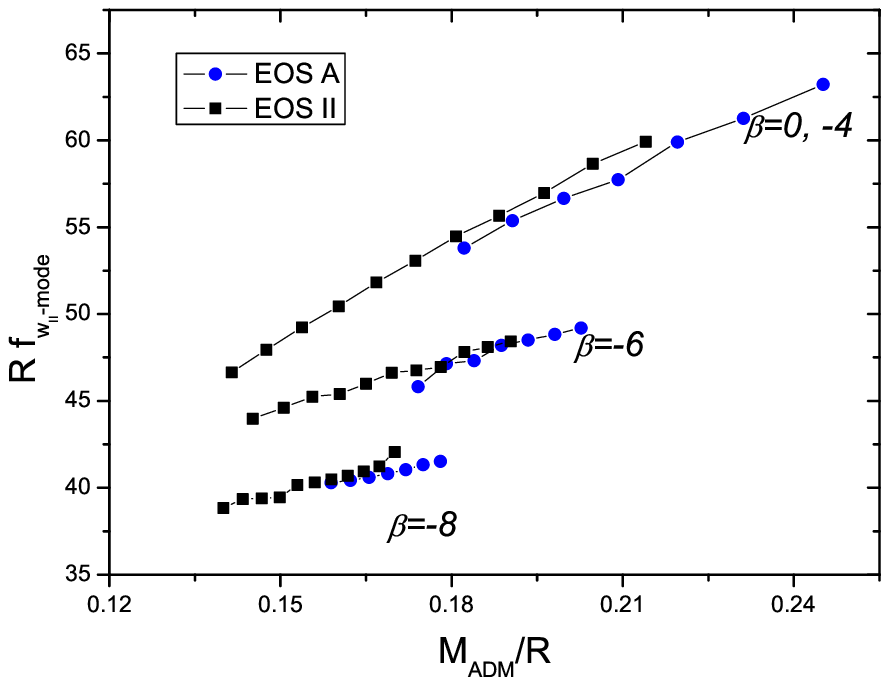}
\caption{The $w_1$ modes (left panel) and the $w_{\rm II}$ modes
(right panel) for EOS A and II and for values of $\beta=0$, -4, -6
and -8.}
\label{fig_RewRvsMR}
\end{figure}
\end{center}

\begin{center}
\begin{figure}[htbp]
\includegraphics[height=6cm]{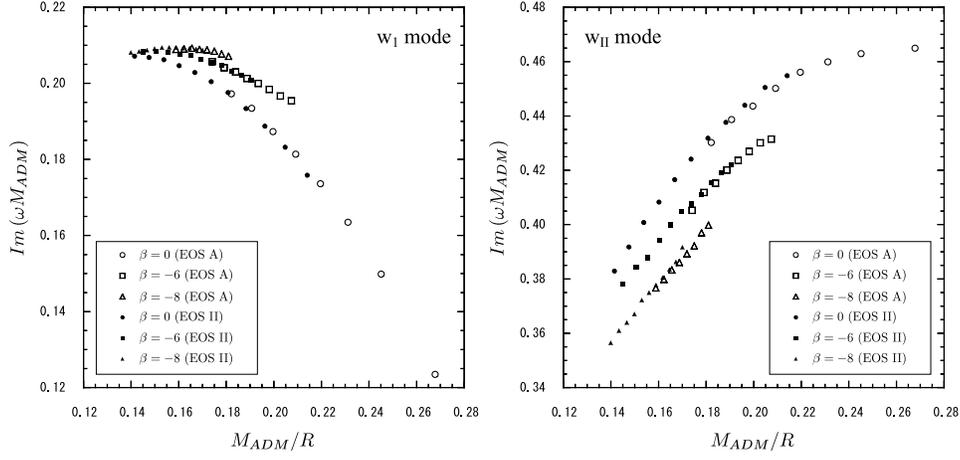}
\caption{ Dependence of the imaginary part of $w_1$ and $w_{\rm II}$
mode on the stellar compactness $M_{ADM}/R$. The left panel is for
the $w_1$ mode and the right for the $w_{\rm II}$ mode. It is
apparent that the imaginary part for $w_{\rm II}$ mode can decrease
by up to 30\% in the presence of a scalar field. For the $w_1$
modes, on the contrary, the damping time becomes shorter.}
\label{fig_w1}
\end{figure}
\end{center}

The results of the two methods we described agree very well, providing
a good consistency check on our calculations.  In Figures
\ref{fig_m140}, \ref{fig_RewRvsMR} and \ref{fig_w1}, we present the
eigenvalues. Our results suggest that the presence of a spontaneous
scalarization can be inferred from the $w$ modes emitted by a newly
born, oscillating neutron star.

In Figure \ref{fig_m140} we show the eigenfrequencies of the $w$
modes for neutron star models with $M_{ADM}=1.4M_{\odot}$. The plot
is reminiscent of earlier calculation of these modes (see e.g.
Figure 3 in \cite{Kokkotas1999}). The modes that might be relevant
for gravitatonal wave detectors are the lowest $w$ modes
\cite{Kokkotas1992}.  The $w_{\rm II}$ modes \cite{Nollert1993} damp out roughly twice
as fast as the $w$ modes, but having lower frequencies they could
also be relevant for detection by Earth-based interferometers. The
higher-frequency $w$ modes  ($w_2$, $w_3$, $w_4$, ...)are difficult,
if not impossible to detect.

In the study of $w$ modes as a tool for asteroseismology
\cite{Andersson1996,Andersson1998,Benhar1999,Kokkotas2001} it has
been suggested that a proper normalization for $Re(\omega)$ is to
multiply it with the radius $R$ of the star and to scale it as a
function of the compactness $M/R$. This phenomenological argument
has been recently verified analytically by Tsui and Leung
\cite{Tsui2005}. Introducing $f=Re(\omega)/2\pi$, it is clear that
$R f$ scales linearly as function of the compactness $M/R$. This
applies both to $w_{\rm II}$ and $w_1$ modes (and even to the
higher overtones). The linear relations that can be derived from
Figure \ref{fig_RewRvsMR} are
\begin{equation}
 f_{\omega_1{\rm -mode}} (\mbox{kHz}) =\frac{1}{{\bar R}}\left( \alpha_1-\beta_1
    \frac{{\bar M}}{{\bar R}} \right) \quad \mbox {and} \quad
 f_{\omega_{\rm II}{\rm -mode}} (\mbox{kHz})
   =\frac{1}{{\bar R}}\left( \alpha_{\rm II}+\beta_{\rm II} \frac{{\bar
   M}}{{\bar R}} \right)\,, \label{eq:w-fittng}
\end{equation}
where the constants $\alpha_1$, $\beta_1$, $\alpha_{\rm II}$ and
$\beta_{\rm II}$ are listed in Table \ref{Tab:fitting}.
\begin{table}
\caption{The coefficients for the fitting factors of equations
(\ref{eq:w-fittng}).} \label{Tab:fitting}
\begin{tabular}{cc|ccccc}
  \hline
  \hline
  & $\beta$ & $\alpha_1$ & $\beta_1$ & $\alpha_{\rm II}$ & $\beta_{\rm II}$ & \\
  \hline
  &  0 & 13.35 & 4.20 & 2.48 & 3.32 & \\
  & -6 & 13.57 & 4.89 & 3.09 & 1.88 & \\
  & -8 & 13.36 & 5.17 & 2.89 & 1.50 & \\
  \hline
  \hline
\end{tabular}
\end{table}

Another reason why it is harder to detect high-damped quasinormal
modes such as the $w_{\rm II}$ modes for compact stars is that the
effective amplitude scales as the square root of the number of
oscillations \cite{Kokkotas2001}.  Typically we can hardly observe
more than $2-3$ cycles for highly damped quasinormal modes of
black-holes and for the $w_{\rm II}$ modes of compact stars.
Spontaneous scalarization might help in this direction. Figure
\ref{fig_w1} shows that the damping time of the $w_{\rm II}$ mode
for stars with $\beta \alt -4.35$ is significantly longer than for
typical stars in general relativity. The reason is that the presence
of a scalar field increases the maximum mass of the stars and their
compactness. Since the damping scales with compactness, the $w_{\rm
II}$ modes live considerably longer.  On the contrary, the damping
times of the $w_1$ modes become shorter as the compactness increases
(left panel in Figure \ref{fig_w1}).

\section{Conclusion}
\label{sec:V}

We derived the equations describing stellar perturbations in
scalar-tensor theories of gravity. The presence of a scalar field
affects the equilibrium model, and consequently the oscillation
spectrum. The scalar field perturbations couple with the polar
perturbations of the spacetime and fluid, but they don't couple with
the axial perturbations. Since the spacetime modes of polar and axial
perturbations have the same qualitative behavior, we have chosen to
study the effect of the scalar field on the axial perturbations.

The results show that in the presence of spontaneous scalarization,
a scalar field reduces the oscillation frequency of the $w_1$ modes
by about 10\% (i.e. by about 1kHz). The decrease in frequency for
the $w_{\rm II}$ modes is about 25\% the frequency of (i.e., about
1.5 kHz). The effect on the damping time is even more pronounced.
The damping of $w_{\rm II}$ modes can decrease by as much as 30\%, while it
can increase by as much as 50\% for the $w_1$ modes.
Detectors operating at these high frequencies are under development.
Through a detection of the $w$ mode spectrum, they could provide a
unique proof for the existence of scalar fields with $\beta \alt
-4.35$.

A more detailed model of the effect of the scalar field on the
oscillation spectra requires the inclusion of a larger set of
equations of state. Another open problem is the study of polar
oscillations, which couple directly to the scalar field.  Work in
these directions is in progress.

%
%


\acknowledgments

We acknowledge valuable comments by Emanuele Berti and Shijun
Yoshida. This work was partially supported by a Grant for The 21st
Century COE Program (Holistic Research and Education Center for
Physics Self-Organization Systems) at Waseda University and the
Pythagoras I research grant of GSRT.

\appendix

\section{The Perturbed Energy Momentum Tensor}          
\label{sec:appendix_2}

In this Appendix we show the explicit form of the various components
of the perturbed energy momentum tensor (of the fluid and of the
scalar field) appearing in the perturbation equations. We will use
primes for spatial derivatives and dots for temporal derivatives.
For simplicity we will omit the subscript $lm$ in the various
perturbed quantities.

The components of the perturbed energy momentum tensor
$T_{*\mu\nu}^{(\varphi)}$ for the scalar field have the form
\begin{align}
 \delta T_{*tt}^{(\varphi)} &= e^{2\Phi - 2\Lambda}\left[2\Psi \delta \varphi'
                             - \left(\hat{H}_0 + \hat{H_2}\right) \Psi^2\right] Y_{lm}, \\
 \delta T_{*tr}^{(\varphi)} &= \left[2\Psi \delta \dot{\varphi} - e^{-2\Lambda} \Psi^2 \hat{H}_1\right] Y_{lm}, \\
 \delta T_{*t \theta}^{(\varphi)} &= e^{-2\Lambda} \Psi^2 \hat{h}_0
                                     \frac{1}{\sin \theta} \partial_{\phi} Y_{lm}, \\
 \delta T_{*t \phi}^{(\varphi)}   &= -e^{-2\Lambda} \Psi^2 \hat{h}_0
                                     \sin \theta \partial_{\theta} Y_{lm}, \\
 \delta T_{*rr}^{(\varphi)} &= 2\Psi \delta \varphi' Y_{lm}, \\
 \delta T_{*r \theta}^{(\varphi)} &= \left[2\Psi \delta \varphi \partial_{\theta}
                                   + e^{-2\Lambda} \Psi^2 \hat{h}_1
                                     \frac{1}{\sin \theta} \partial_{\phi}\right] Y_{lm}, \\
 \delta T_{*r \phi}^{(\varphi)}   &= \left[2\Psi \delta \varphi \partial_{\phi}
                                   - e^{-2\Lambda} \Psi^2 \hat{h}_1
                                     \sin \theta \partial_{\theta}\right] Y_{lm}, \\
 \delta T_{*\theta \theta}^{(\varphi)} &= r^2 e^{-2\Lambda}\left[ -2\Psi \delta \varphi'
                                        + \left(\hat{H}_2 - \hat{K}\right) \Psi^2 \right] Y_{lm}, \\
 \delta T_{*\phi \phi}^{(\varphi)}   &= r^2 e^{-2\Lambda} \left[ -2\Psi \delta \varphi'
                                      + \left(\hat{H}_2 - \hat{K}\right) \Psi^2 \right] \sin^2 \theta Y_{lm}.
\end{align}
In order to get the components of the perturbed energy momentum
tensor for the fluid we define, in the physical frame, the
variations of pressure $\delta \tilde{P}= \delta \tilde{P}\,
Y_{lm}$, energy density $\delta \tilde{\rho}  = \delta
\tilde{\rho}\, Y_{lm}$
and the components of the perturbed 4-velocity (in the physical
frame)
\begin{align}
 \delta \tilde{U}^{t} &= \frac{1}{2A^3} e^{-\Phi} H_0 Y_{lm}, \\
 \delta \tilde{U}^{r} &= \frac{1}{A} e^{\Phi - 2\Lambda}WY_{lm}, \\
 \delta \tilde{U}^{\theta} &= \frac{1}{Ar^2} e^{\Phi} \left(V\partial_{\theta} Y_{lm}
                      - u \frac{1}{\sin \theta} \partial_{\phi} Y_{lm}\right), \\
 \delta \tilde{U}^{\phi}   &= \frac{1}{Ar^2 \sin^2 \theta} e^{\Phi}
                      \left(V\partial_{\phi} Y_{lm} + u \sin \theta \partial_{\theta}Y_{lm}\right),
\end{align}
where the perturbation functions $\delta \tilde{P}$, $\delta
\tilde{\rho}$, $W$, $V$, and $u$ defined in the previous relations
depend only on $t$ and $r$.
Using the above definition  the
components of the perturbed energy-momentum tensor $\delta T_{*\mu\nu}$ take the form
\begin{align}
 \delta T_{*tt} &= A^4 e^{2\Phi} \left[\delta \tilde{\rho} - \tilde{\rho} \hat{H}_0
                 + \frac{4 \tilde{\rho}}{A} \delta A\right] Y_{lm}, \\
 \delta T_{*tr} &=  -A^4 e^{2\Phi} \left[\left(\tilde{\rho} + \tilde{P}\right)W
                    + e^{-2\Phi} \tilde{\rho} \hat{H}_1\right] Y_{lm}, \\
 \delta T_{*t \theta} &= -A^4 e^{2\Phi} \left(\tilde{\rho} + \tilde{P}\right)V\partial_{\theta} Y_{lm}
                       + A^4 e^{2\Phi}\left[\left(\tilde{\rho} + \tilde{P}\right) u
                       + e^{-2\Phi} \tilde{\rho} \hat{h}_0\right]
                         \frac{1}{\sin \theta} \partial_{\phi} Y_{lm}, \\
 \delta T_{*t \phi}   &= -A^4 e^{2\Phi} \left(\tilde{\rho} + \tilde{P}\right)V\partial_{\phi} Y_{lm}
                       - A^4 e^{2\Phi} \left[\left(\tilde{\rho} + \tilde{P}\right) u
                       + e^{-2\Phi} \tilde{\rho} \hat{h}_0\right]
                         \sin \theta \partial_{\theta} Y_{lm}, \\
 \delta T_{*rr} &= A^4 e^{2\Lambda} \left[\delta \tilde{P} + \tilde{P} \hat{H}_2
                 + \frac{4\tilde{P}}{A} \delta A\right]Y_{lm}, \\
 \delta T_{*r \theta} &= -A^4 \tilde{P} \hat{h}_1 \frac{1}{\sin \theta} \partial_{\phi} Y_{lm}, \\
 \delta T_{*r \phi}   &= A^4 \tilde{P} \hat{h}_1 \sin \theta \partial_{\theta} Y_{lm}, \\
 \delta T_{*\theta \theta} &= A^4 r^2 \left[\delta \tilde{P} + \tilde{P} \hat{K}
                                  + \frac{4\tilde{P}}{A} \delta A\right]Y_{lm}, \\
 \delta T_{*\phi \phi}     &= A^4 r^2 \left[\delta \tilde{P} + \tilde{P} \hat{K}
                              + \frac{4\tilde{P}}{A} \delta A\right] \sin^2 \theta Y_{lm}.
\end{align}

\section{The components of the Linearized Einstein equations}   
\label{sec:appendix_3}

Here we provide the explicit form of the various expressions used in
equations equations (\ref{perturbed-einstein1}) --
(\ref{perturbed-einstein5}) for the description of the perturbed
Einstein equations (\ref{var_basic1}). We have chosen to use the same
notation as Kojima \cite{Kojima1992} to facilitate comparison.
\begin{align}
 A_{lm}^{(0)} =& -\hat{K}'' + e^{2\Lambda}\left(\frac{1}{2}re^{-2\Lambda}\Psi^2 + \frac{5\mu}{r^2} - \frac{3}{r}
                 + 4\pi G_* r A^4\tilde{\rho}\right) \hat{K}'
                 + \frac{1}{r}\hat{H}_2' + \frac{(l-1)(l+2)}{2r^2}e^{2\Lambda} \hat{K}
                 \nonumber \\
                 &+ e^{2\Lambda}\left(\frac{l(l+1)}{2r^2} + \frac{1}{r^2} -8\pi G_*A^4\tilde{\rho}\right)\hat{H}_2
                 - 2\Psi \delta \varphi'
                 -8\pi G_* A^4 e^{2\Lambda} (\delta \tilde{\rho}  + 4\tilde{\rho}\alpha \delta \varphi), \label{Alpha0_lm}\\
 A_{lm}^{(1)} =& -\dot{\hat{K}}
                 + e^{2\Lambda}\left(\frac{1}{2}re^{-2\Lambda}\Psi^2 + \frac{3\mu}{r^2} - \frac{1}{r}
                 + 4\pi G_* r A^4 \tilde{P} \right)\dot{\hat{K}}
                 + \frac{1}{r}\dot{\hat{H}}_2
                 + \frac{l(l+1)}{2r^2} \hat{H}_1 \nonumber \\
                 &- 2\Psi \delta \dot{\varphi} + 8\pi G_* A^4 e^{2\Phi} \left(\tilde{\rho} + \tilde{P}\right)W, \label{Alpha1_lm}\\
 A_{lm}^{(2)} =& -e^{-2\Phi + 2\Lambda} \ddot{{\hat{K}}}
                + e^{2\Lambda}\left(\frac{1}{2}re^{-2\Lambda}\Psi^2 - \frac{\mu}{r^2} + \frac{1}{r}
                + 4\pi G_* r A^4 \tilde{P}\right) \hat{K}' + \frac{2}{r} e^{-2\Phi} \dot{\hat{H}}_1
                - \frac{1}{r} \hat{H}_0' - \frac{(l-1)(l+2)}{2r^2} e^{2\Lambda} \hat{K} \nonumber \\
                &+ \frac{l(l+1)}{2r^2} e^{2\Lambda} \hat{H}_0
                - e^{2\Lambda} \left(\frac{1}{r^2} + 8\pi G_* A^4 \tilde{P}\right) \hat{H}_2
                - 2\Psi \delta \varphi'
                - 8\pi G_* A^4 e^{2\Lambda} \left(\delta \tilde{P} + 4\tilde{P}\alpha \delta \varphi\right), \label{Alpha2_lm}
\end{align}
 \begin{align}
 A_{lm}^{(3)} =& \hat{K}'' - \hat{H}_0''
           - e^{-2\Phi + 2\Lambda} \left(\ddot{\hat{K}} + \ddot{\hat{H}}_2\right)
           + 2 e^{-2\Phi} \dot{\hat{H}}_1'
           - e^{2\Lambda}\left(\frac{1}{2}re^{-2\Lambda}\Psi^2 + \frac{r+\mu}{r^2}
           + 4\pi G_* \left(2\tilde{P} - \tilde{\rho}\right)r A^4\right) \hat{H}_0' \nonumber \\
           &- e^{2\Lambda}\left(\frac{1}{2}re^{-2\Lambda}\Psi^2 - \frac{\mu}{r^2} + \frac{1}{r}
           + 4\pi G_* r A^4 \tilde{P}\right) \hat{H}_2'
           + e^{2\Lambda}\left(4\pi G_* \left(\tilde{P} - \tilde{\rho}\right)rA^4
           + \frac{2(r-\mu)}{r^2}\right) \hat{K}'  \nonumber \\
           &- 2 e^{-2\Phi + 2\Lambda} \left(\frac{1}{2}re^{-2\Lambda}\Psi^2 + \frac{\mu-r}{r^2}
           + 4\pi G_* r A^4 \tilde{\rho}\right)\dot{\hat{H}}_1
           + \frac{l(l+1)}{2r^2} e^{2\Lambda} \hat{H}_0 \nonumber \\
           &- e^{2\Lambda} \left(16\pi G_*A^4 \tilde{P} + \frac{l(l+1)}{2r^2}\right) \hat{H}_2
           + 4\Psi \delta \varphi'
           -16\pi G_* A^4 e^{2\Lambda} \left(\delta \tilde{P} + 4\tilde{P}\alpha \delta \varphi\right), \label{Alpha3_lm}
 \end{align}
 \begin{align}
 \alpha_{lm}^{(0)} =& \frac{1}{2} e^{-2\Lambda}\left[\hat{H}_1'
                      - e^{2\Lambda}\left(\dot{\hat{H}}_2 + \dot{\hat{K}}\right)
                      + e^{2\Lambda}\left(4\pi G_* \left(\tilde{P} - \tilde{\rho}\right) r A^4
                      + \frac{2\mu}{r^2}\right)\hat{H}_1\right]
                      + 8\pi G_* A^4 e^{2\Phi} \left(\tilde{\rho} + \tilde{P}\right)V, \label{alpha0_lm} \\
 \alpha_{lm}^{(1)} =& \frac{1}{2} \Bigg[\hat{H}_0' - \hat{K}' -e^{-2\Phi} \dot{\hat{H}}_1
                      + e^{2\Lambda}\left(\frac{1}{2}re^{-2\Lambda}\Psi^2 + \frac{3\mu}{r^2} - \frac{1}{r}
                      + 4\pi G_* r A^4 \tilde{P} \right) \hat{H}_0 \nonumber \\
                      &+ e^{2\Lambda}\left(\frac{1}{2}re^{-2\Lambda}\Psi^2 - \frac{\mu}{r^2} + \frac{1}{r}
                      + 4\pi G_* r A^4 \tilde{P}\right) \hat{H}_2\Bigg] - 2\Psi \delta \varphi, \label{alpha1_lm}\\
 \beta_{lm}^{(0)} =& \frac{1}{2}e^{-2\Lambda} \left(\hat{h}_0'' - \dot{\hat{h}}_1'\right)
                      - \left\{2\pi G_* \left(\tilde{\rho} + \tilde{P}\right)rA^4
                      + \left(\frac{r}{2}-\mu\right)\Psi^2\right\} \left(\hat{h}_0' - \dot{\hat{h}}_1\right)
                      - \frac{1}{r} e^{-2\Lambda} \dot{\hat{h}}_1 \nonumber \\
                      &- \frac{1}{2r^3} \left\{l(l+1)r - 4\mu
                      + 8 \pi G_* \left(\tilde{\rho} + \tilde{P}\right) r^3A^4
                      - 2r^3 e^{-2\Lambda} \Psi^2 \right\} \hat{h}_0
                      - 8\pi G_* A^4 e^{2\Phi} \left(\tilde{\rho} + \tilde{P}\right)u, \label{beta0_lm}\\
 \beta_{lm}^{(1)} =& \frac{1}{2} e^{-2\Phi} \left(\dot{\hat{h}}_0' - \ddot{\hat{h}}_1\right)
                      - \frac{1}{r} e^{-2\Phi} \dot{\hat{h}}_0
                      - \frac{(l-1)(l+2)}{2r^2} \hat{h}_1, \label{beta1_lm}\\
 s_{lm} =& \frac{1}{2} \left(\hat{H}_0 - \hat{H}_2\right), \label{s_lm} \\
 t_{lm} =& e^{-2\Phi} \dot{\hat{h}}_0 - e^{-2\Lambda} \hat{h}_1'
                           - \frac{1}{r^2} \left\{2\mu
                           + 4\pi G_* \left(\tilde{P}-\tilde{\rho}\right)r^3 A^4\right\}\hat{h}_1, \label{t_lm}
\end{align}
where the equations are simplified by virtue of the equations (13) -- (17) in Paper I.

\section{Numerical techniques}          
\label{sec:appendix_4}

In this Appendix we present two numerical techniques to determine
quasinormal modes. The first is the direct evolution of the time
dependent, axial perturbation equation (\ref{axial_wave}).  In the
second approach we assume a harmonic decomposition for the
perturbation function $X$ of the form $X(r,t)=X(r)e^{i\omega t}$, and
solve the equation (\ref{wave-equation}) as the eigenvalue problem.
We shall consider the equations (\ref{wave-equation})
in the interior and the exterior
of the star; then we will find the eigenvalues (quasinormal modes) by
matching the interior and exterior solutions.

The largest numerical error in the interior solution occurs at
stellar surface, where the pressure is zero. In order to avoid this
difficulty, we integrate the perturbation equation
(\ref{wave-equation}) from both sides, i.e. from the stellar center
$r=0$ and from the stellar surface $r=R$. Then we match the
solutions at some intermediate point, e.g., $r=R/2$ (see for example
\cite{LD83,Kokkotas1992}). In order to deal with the boundary
condition at infinity we adopt the continued fraction method,
originally used for black hole perturbations by Leaver
\cite{Leaver1985}. To use this method we must know the forms of the
coefficient in the perturbation equation as functions of $1/r$.
Because of the presence of a scalar field , we do not know the exact
forms of these coefficients. Therefore we just use the asymptotic
forms of the coefficients, and derive a five-term recurrence
relation. We believe that the QNMs obtained using these asymptotic
forms are accurate enough, because the difference between the value
of $\mu$ at the stellar surface and at infinity is not so large.

\subsection{Interior region of the star}

The numerical integration of Equation (\ref{wave-equation}) inside
the star will be split (for numerical reasons) into two parts. First,
we will integrate  Equation (\ref{wave-equation}) from the center
towards $R/2$ and then we will integrate from the surface towards
the same point. The matching of the two solutions will provide a
unique solution valid throughout the star.

Near the center it can be shown that $X$ has a behavior  of the
form
\begin{align}
 X = X_c r^{l+1}\left(1 + O(r^{2}) \right), \label{BC_center}
\end{align}
where $X_c$ is some arbitrary constant. Using this boundary
condition (\ref{BC_center}) and by integrating equation
(\ref{wave-equation}) from $r=0$ to the matching point $r=R/2$, one
can obtain the values of  $X(r)$ and $X'(r)$. For convenience we
represent the two functions $X$ and $X'$  in the vector form ${\bf
Y}=\left(X,X'\right)$ and we will call ${\bf Y}_{0}(r)$ the solution
in the range $0\le r\le R/2$. The next step will be to integrate
equation (\ref{wave-equation}) from the stellar surface towards
$R/2$ with a set of boundary conditions at $r=R$ such as
$(X(R),X'(R))=(1,0)$ and $(X(R),X'(R))=(0,1)$. In this way we get
two independent solutions ${\bf Y}_1(r)$ and ${\bf Y}_2(r)$
corresponding to each one of the previous boundary conditions. Thus
the solution of the perturbation equation (\ref{wave-equation}) is
\begin{align}
 {\bf Y}(r) &= {\bf Y}_0(r),\ \ {\mbox{for}}\ 0\le r \le R/2, \\
 {\bf Y}(r) &= a{\bf Y}_1(r) + b{\bf Y}_2(r), \ \ {\mbox{for}}\ R/2\le r \le R,
\end{align}
where $a$ and $b$ are some constant, which will be determined from
the junction condition at $r=R/2$:
\begin{equation}
 {\bf Y}_0(R/2) = a{\bf Y}_1(R/2) + b{\bf Y}_2(R/2).
\end{equation}
The determination of the two constants specifies uniquely the
solution in the interior of the star for a given value of the
frequency $\omega$ and the constant $X_c$.  At the stellar surface
the values of $X(R)$, $X'(R)$ are simply $X(R)= a$ and $X'(R)= b$.

\subsection{Exterior region of the star}

The functions describing the stellar background simplify
considerably outside the star.  This leads to a corresponding
simplification of the wave equation (\ref{wave-equation}).  In the
exterior, the equations describing the background reduces to
\begin{align}
 \mu '     &= \frac{1}{2} r e^{-2\Lambda} \Psi^2, \label{outer_dmu} \\
 \Phi '    &= \frac{1}{2}r\Psi^2 + \frac{\mu}{r^2} e^{2\Lambda}, \\
 \varphi ' &= \Psi, \\
 \Psi '    &= -\frac{2}{r^2}(r-\mu)e^{2\Lambda}\Psi. \label{outer_dPsi}
\end{align}
Therefore the asymptotic form of the above background quantities are
\begin{align}
 \mu     &= M_{ADM} + \frac{\mu_1}{r} + O\left(\frac{1}{r^2}\right), \label{mu1} \\
 \Phi    &= -\frac{M_{ADM}}{r} + O\left(\frac{1}{r^2}\right), \label{Phi1} \\
 \varphi &= \varphi_0 + \frac{\omega_A}{r} + O\left(\frac{1}{r^2}\right),
\end{align}
where $\mu_1= -\omega_A^2/2$ and $\omega_A= -M_{\mbox{\small ADM}}\Psi_s/\Phi'_s$.

The perturbation equation (\ref{wave-equation}) in view of the above
relations, outside the star, get the form
\begin{align}
 \left(1-\frac{2\mu}{r} \right) X'' + \frac{2\mu}{r^2} X'
     + \left[\omega^2 e^{-2\Phi} - \frac{l(l+1)}{r^2} + \frac{6\mu}{r^3} \right] X = 0. \label{wave-equation1}
\end{align}
which is similar (in the absence of a scalar field, identical) to
the Regge-Wheeler \cite{RW1957} equation describing the axial
perturbations in the exterior of a spherically symmetric spacetime
(either a black-hole or a neutron star). Using as boundary values
for the integration the values of $X$ and $X'$ at the surface given
by the two relations $X(R)= a$ and $X'(R)= b$
one can integrate equation (\ref{wave-equation1}) together with
(\ref{outer_dmu}) -- (\ref{outer_dPsi}) from the stellar surface
towards infinite. The numerical integration will obviously stop at
some distance $r=r_a$, where we will have to match the numerical
solution with the appropriate asymptotic boundary conditions (in
this case, the absence of incoming radiation).

In order to find the asymptotic form of the solution of  equation
(\ref{wave-equation1}) we can assume a solution of the form
\begin{equation}
 X(r) = \left(\frac{r}{2\hat{M}} -1 \right)^{-2i\hat{M}\omega} e^{-i\omega r}
       \sum_{n=0}^{\infty} a_n \left(1-\frac{r_a}{r}\right)^n,
\end{equation}
where $\hat{M}=M_{ADM}$. By substituting this form of the solution
into the perturbation equation (\ref{wave-equation1}) and taking the
leading orders for $\mu$ and $\Phi$, i.e., keeping only the terms up
to order $1/r$, from equations (\ref{mu1}) and (\ref{Phi1}), we
obtain a five-term recurrence relation for the expansion
coefficients $a_n$ $(n\ge 1)$,
\begin{equation}
 \alpha_n a_{n+1} + \beta_n a_n + \gamma_n a_{n-1} + \delta_n a_{n-2}
 + \epsilon_n a_{n-3} = 0, \label{recurrence}
\end{equation}
where the coefficients of the recurrence relation are given by the
following formulae
\begin{align}
 \alpha_n   &= c_0 n (n+1), \\
 \beta_n    &= d_0 n + c_1 n (n-1), \\
 \gamma_n   &= e_0 + d_1 (n-1) + c_2 (n-1)(n-2), \\
 \delta_n   &= e_1 + d_2 (n-2) + c_3 (n-2)(n-3), \\
 \epsilon_n &= e_2 + d_3 (n-3) + c_4 (n-3)(n-4).
\end{align}
The coefficients $c_i$, $d_i$ and $e_i$ are functions of the
background quantities and have the form
\begin{align}
 c_0 &= 1-\frac{2 \hat{M}}{r_a} - \frac{2\mu_1}{{r_a}^2}, \\
 c_1 &= -2 + \frac{6 \hat{M}}{r_a} + \frac{8\mu_1}{{r_a}^2}, \\
 c_2 &= 1 - \frac{6 \hat{M}}{r_a} - \frac{12 \mu_1}{{r_a}^2}, \\
 c_3 &= \frac{2 \hat{M}}{r_a} + \frac{8\mu_1}{{r_a}^2}, \\
 c_4 &= -\frac{2\mu_1}{{r_a}^2},
\\
 d_0 &= -2i\omega r_a -2 + \frac{6\hat{M}}{r_a} + \frac{4i\omega \mu_1}{r_a}
      + \frac{8i\omega \hat{M}\mu_1}{{r_a}^2} + \frac{6\mu_1}{{r_a}^2}, \\
 d_1 &= 2 - \frac{12\hat{M}}{r_a} - \frac{8i\omega \mu_1}{r_a} - \frac{24i\omega \hat{M}\mu_1}{{r_a}^2}
      - \frac{18\mu_1}{{r_a}^2}, \\
 d_2 &= \frac{2}{r_a}\left( 3\hat{M} + 2i\omega \mu_1 + \frac{12i\omega \hat{M}\mu_1}{r_a} + \frac{9\mu_1}{r_a}\right), \\
 d_3 &= -\frac{2\mu_1}{{r_a}^2} \left(3 + 4i\omega \hat{M}\right),
\end{align}
 \begin{align}
 e_0 &= -l(l+1) + 2\omega^2 \mu_1 + \frac{6\hat{M}}{r_a} + \frac{8\omega^2 \hat{M}\mu_1}{r_a} - \frac{2i\omega \mu_1}{r_a}
      - \frac{8i\omega \hat{M}\mu_1}{{r_a}^2} + \frac{6\mu_1}{{r_a}^2}, \\
 e_1 &= \frac{2}{r_a}\left(-3\hat{M} - 4\omega^2 \hat{M}\mu_1 + i\omega \mu_1 + \frac{8i\omega \hat{M}\mu_1}{r_a}
      - \frac{6\mu_1}{r_a}\right), \\
 e_2 &= \frac{2\mu_1}{{r_a}^2}\left(3 - 4i\omega \hat{M}\right).
\end{align}
The first four terms of the recurrence relation (\ref{recurrence})
$a_{-2}$, $a_{-1}$, $a_0$, and $a_1$ are provided by the values of
$X$ and $X'$ at $r=r_a$, i.e.,
\begin{equation}
a_{-2}= a_{-1} = 0,\qquad a_0    = \frac{X(r_a)}{\Xi(r_a)},\quad
\mbox{and} \quad
 a_1    = \frac{r_a}{\Xi(r_a)} \left[X'(r_a) + \frac{i\omega r_a}{r_a - 2\hat{M}} X(r_a)\right],
\end{equation}
where
\begin{equation}
 \Xi(r) = \left(\frac{r}{2\hat{M}} - 1 \right)^{-2i\hat{M}\omega} e^{-i\omega r}.
\end{equation}
The five term recurrence relations have in principle four possible
solutions. A high order recurrence relation can generally be reduced
to a three term recurrence relation, in which case convergence
criteria for the solution can be easily applied, and we can identify
the solution describing only outgoing radiation \cite{Leaver1985}.
To obtain a three-term recurrence relation we define new
coefficients $\hat{\alpha}_n$, $\hat{\beta}_n$, $\hat{\gamma}_n$,
and $\hat{\delta}_{n}$ as
\begin{equation}
\hat{\alpha}_n = \alpha_n,\ \ \hat{\beta}_n = \beta_n,\ \
\hat{\gamma}_n = \gamma_n \quad \mbox{for} \quad n=1,2
\end{equation}
and for $n\ge 3$
\begin{equation}
 \hat{\alpha}_n = \alpha_n, \quad
 \hat{\beta}_n  = \beta_n - \frac{\hat{\alpha}_{n-1} \epsilon_n}{\hat{\delta}_{n-1}},
 \quad
 \hat{\gamma}_n = \gamma_n - \frac{\hat{\beta}_{n-1} \epsilon_n}{\hat{\delta}_{n-1}},
 \quad
 \hat{\delta}_n = \delta_n - \frac{\hat{\gamma}_{n-1}
 \epsilon_n}{\hat{\delta}_{n-1}}\, .
\end{equation}
The original five-term recurrence relation becomes a four-term
recurrence relation:
\begin{equation}
 \hat{\alpha}_n a_{n+1} + \hat{\beta}_n a_n + \hat{\gamma}_n a_{n-1} + \hat{\delta}_n a_{n-2} = 0. \label{four-term}
\end{equation}
Note that for the case of $\mu_1=0$, that is the case of the
standard neutron star obtained by Einstein's theory for gravity
($\beta=0$), the recurrence relation for $a_n$ has four terms
\cite{Benhar1999,Sotani2001,Berti2004}.

The final step will be to define another set of coefficients
$\tilde{\alpha}_n$, $\tilde{\beta}_n$, and $\tilde{\gamma}_n$:
\begin{equation}
 \tilde{\alpha}_1 = \hat{\alpha}_1,\ \ \tilde{\beta}_1 = \hat{\beta}_1,\ \ \tilde{\gamma}_1 = \hat{\gamma}_1,
\end{equation}
and for $n\ge 2$
\begin{equation}
 \tilde{\alpha}_n = \hat{\alpha}_n, \quad
 \tilde{\beta}_n  = \hat{\beta}_n - \frac{\tilde{\alpha}_{n-1} \hat{\delta}_n}{\tilde{\gamma}_{n-1}}
 \quad \mbox{and} \quad
 \tilde{\gamma}_n = \hat{\gamma}_n - \frac{\tilde{\beta}_{n-1} \hat{\delta}_n}{\tilde{\gamma}_{n-1}},
\end{equation}
The four-term recurrence relation (\ref{four-term}) is thus reduced to
a three-term relation of the form
\begin{align}
 \tilde{\alpha}_n a_{n+1} + \tilde{\beta} a_n + \tilde{\gamma}_n a_{n-1} = 0.
\end{align}
Using this three-term recurrence relation, the boundary condition can
be expressed as a continued fraction relation between
$\tilde{\alpha}_n$, $\tilde{\beta}_n$, and $\tilde{\gamma}_n$:
\begin{equation}
 \frac{a_1}{a_0} = \frac{-\tilde{\gamma}_1}{\tilde{\beta}_1-}
                   \frac{\tilde{\alpha}_1\tilde{\gamma}_2}{\tilde{\beta}_2-}
                   \frac{\tilde{\alpha}_2\tilde{\gamma}_3}{\tilde{\beta}_3-} \dots,
\end{equation}
that can be rewritten as
\begin{equation}
 0 = \tilde{\beta}_0 -
     \frac{\tilde{\alpha}_0\tilde{\gamma}_1}{\tilde{\beta}_1-}
     \frac{\tilde{\alpha}_1\tilde{\gamma}_2}{\tilde{\beta}_2-}
     \frac{\tilde{\alpha}_2\tilde{\gamma}_3}{\tilde{\beta}_3-} \dots \equiv f(\omega),
\end{equation}
where $\tilde{\beta}_0 \equiv a_1/a_0$, $\tilde{\alpha}_0\equiv -1$.
The eigenfrequency $\omega$ of a quasinormal mode can be obtained
solving the equation $f(\omega)=0$.


\end{document}